\documentclass[12pt,a4paper]{article}
\usepackage{amsmath,amssymb}
\usepackage{exscale}
\usepackage{relsize}
\usepackage{graphicx}

\def\endproof{\par\nobreak\hbox to \hsize{\hfil\vrule width 5pt height
5pt}\goodbreak\vskip 3pt}

\begin{document}

\title{Thermodynamical identities---a systematic approach}
\author{J. B. Cooper\\ Johannes Kepler Universit\"at Linz}
\date{}
\maketitle

\begin{abstract}
We present a systematic
approach to thermodynamical identities and illustrate the power of these
methods by displaying Mathematica notebooks  
to deal with a large variety of such identities.
In concrete examples these
can  involve
rather tedious, not to say impossible,  computations when done by hand.

\end{abstract}

\section{Introduction}
The subject of thermodynamics is notoriously difficult for mathematicians.
V.I. Arnold [Ar] famously put it in a nutshell as follows:
 
\begin{quote}
Every mathematician knows that it is impossible to understand 
any elementary
course in thermodynamics.  
\end{quote}
He continues by explaining that

\begin{quote}
the reason is that [the] thermodynamics is based on a rather complicated 
mathematical theory, on [the] contact geometry.
\end{quote}

In [Co] we presented an axiomatic approach to Gibbsian thermodynamics
which avoided these difficulties and developed a systematic approach to the 
classical identities of
thermodynamics.  
In this short article we wish to present a theory-free version of the latter
topic with a view to
making it directly accessible to readers without the theoretical 
framework  presented in [Co].

The four
Maxwell relations in thermodynamics, which are equivalent to
the Jacobian identity 
$\dfrac {\partial (T,S)}{\partial (p,V)} = 1$,
are our starting point.
Geometrically, they mean that the the corresponding map from the  $(p,V)$-plane
into the $(T,S)$-plane is area-preserving.

\section{The notation}
In order to develop a systematic approach we prefer to employ a mathematically
neutral notation rather than
the standard notation of thermodynamics.
We begin by recalling the latter for the reader's convenience.

Our starting point is the situation where we are 
given the temperature  $T$ and the entropy $S$ as functions of
the pressure $p$ and the volume $V$. 

We use the following dictionary to jump between the purely 
mathematical notation and the thermodynamical one: 
$u$ corresponds to $T$, $v$ to $S$, $p$ to $x$ and $V$ to $y$.  For example,
 the
thermodynamical equations

$$T = p V, \quad S = \frac1{\gamma -1}\ln \left(pV^\gamma\right) $$

\noindent of the ideal gas (we are omitting constants) corresponds to

$$u(x,y) = x y,\quad  v(x,y) = \frac 1 {\gamma -1}\left (\ln x +
\gamma  \ln y \right ).$$

It is a consequence of the Maxwell relations that one can define 
four energy type functions and we will discuss this in more detail
below. 

\section{Thermodynamical identities--- the basic\\ machinery}

We suppose that $u$ and $v$ are given as functions $f$ and $g$ of $x$
and $y$. Thus $u=f(x,y)$, $v=g(x,y)$. When $J=1$, simple manipulations
with differential forms proved the basic identities:
$$\begin{array}{lclclcclclcl} 
du&=&f_1\, dx&+&f_2\, dy,  &&dv&=&g_1\, dx&+&g_2\, dy\\
&&&&&&&&&&\\
dx&=&g_2\, du&-&f_2\, dv, &&dy&=&-g_1\, du &+&f_1\, dv\\
&&&&&&&&&&\\
du&=&\frac{f_2}{g_2}\, dv&+&\fbox{\text{ $\frac 1{g_2}$}}\, dx, 
&&dy&=&\fbox{\text{$ \frac
1{g_2}$}}\, dv&-&\frac{g_1}{g_2}\, dx\\
&&&&&&&&&&\\
du&=&\frac{f_2}{g_1}\, dv&-&\fbox{\text{$ \frac 1{g_1}$}}\, dy, 
&&dx&=&\fbox{\text{$ \frac 1{g_1}$}}\, dv&-&\frac{g_2}{g_1}\, dy\\
&&&&&&&&&&\\
dv&=&\frac{g_1}{f_1}\, du&+&\fbox{\text{$ \frac 1{f_1}$}}\, dy, 
&&dx&=&\fbox{\text{$ \frac1{f_1}$}}\, du&
-&\frac{f_2}{f_1}\, dy\\
&&&&&&&&&&\\
dv&=&\frac{g_2}{f_2}\, du&-&\fbox{\text{$ \frac 1{f_2}$}}\, dx, 
&&dy&=&\fbox{\text{$ \frac
1{f_2}$}}\, du&-&\frac{f_1}{f_2}\, dx\end{array}$$ where  we have 
highlighted the expressions which correspond to the 
Maxwell relations.

\noindent In thermodynamic notation these are
$$\begin{array}{lclcclcl}
dT&=&f_1\, dp+f_2\, dV,&&dS&=&g_1\, dp+g_2\, dV\\
dp&=&g_2\, dT-g_2\, dS,&&dV&=&-g_1\, dT+f_1\, dV\\
dT&=&\dfrac{f_2}{g_2}\, dS+\dfrac 1{g_2}\, dp,
                        &&dV&=&\dfrac 1{g_2}\,
dS-\dfrac{g_1}{g_2}\, dp\\
dT&=&\dfrac{f_2}{g_1}\, dS-\dfrac 1{g_1}\, dV,&&dp&=&\dfrac 1{g_1}\,
dS-\dfrac{g_2}{g_1}\, dV\\
dS&=&\dfrac{g_1}{f_1}\, dT+\dfrac 1{f_1}\, dV,&&dp&=&\dfrac 1{f_1}\,
dT-\dfrac{f_1}{g_2}\, dV\\
dS&=&\dfrac{g_2}{f_2}\, dT-\dfrac 1{f_2}\, dp,&&dV&=&\dfrac 1{f_2}\, 
dT-\dfrac{f_1}{f_2}\, dp
\end{array}$$
where we are using the key: $u\leftrightarrow T$, $v\leftrightarrow
S$, $x\leftrightarrow p$, $y\leftrightarrow V$ introduced above. 

\noindent
In order to isolate the underlying patterns, we now use a numerical
code. Thus
$$\begin{array}{ccccc}
u&\rightarrow&3\leftarrow&T\\
v&\rightarrow&4\leftarrow&S\\
x&\rightarrow&1\leftarrow&p\\
y&\rightarrow&2\leftarrow&V.\end{array}$$

\noindent
Partial derivatives will be denoted by triples in  brackets. $(3,1,2)$, 
for example, denotes $\frac{\partial u}{\partial x}|_y$ 
in the neutral notation, $\frac{\partial T}{\partial p}|_V$
in the thermodynamical one.
In general, $(i,j,k)$ denotes the derivative of the variable $i$, regarded as
a function of the $j$-th and $k$-th variable, with respect to the $j$-th
variable. 

By reading off from the above list, we can express each partial derivative 
of the form $(i,j,k)$ in
terms of $f_1,f_2,g_1,g_2$. For example:
$$
(3,1,2)=f_1,\quad
(3,2,1)=f_2,\quad
(4,1,2)=g_1,\quad
(4,2,1)=g_2;$$

$$
(1,3,4)=g_2,\quad
(2,3,4)=-g_1,\quad
(1,4,3)=-f_2,\quad
(2,4,3)=f_1;$$
 and so on.
\noindent 
We can then express any derivative $(a, b, c)$ in terms of ones of the form
$(d, 1, 2)$ or $(e, 2, 1 )$.  Thus the four derivatives with $x$ and $v$ as 
independent
variables can be computed::

$$\begin{array}{rcr}
(3,4,1)&=&\dfrac{(3,2,1)}{(4,2,1)}\\
(2,4,1)&=&\dfrac 1{(4,2,1)}\\
(3,1,4)&=&\dfrac 1{(4,2,1)}\\
(2,1,4)&=&-\dfrac{(4,1,2)}{(4,2,1)}\end{array}$$

Then, as is standard in thermodynamics, we can introduce four energy functions
$E^{13}$, $E^{14}$, $E^{24}$, $E^{14}$ such that $dE^{24}=u\, dv-x\, dy$, 
$dE^{14}=u\,dv+y\,dx$, $dE^{13}=-v\, du+y\, dx$, $dE^{23}=-v\, du-x\, dy$ 
(the superfixes correspond to the independent variables---thus for 
$E^{13}$ these are  $x$ and $u$ i.e. $1$ and $3$).

We will discuss these in more detail below where the rationale of our notation 
will be explained. In terms of the classical 
notation:
$$\begin{array}{lcrcrl}
dE&=&T\, dS&-&p\, dV &\mbox{(energy)}\\
dF&=&-S\, dT&-&p\, dV &\mbox{(free energy)}\\
dG&=&-S\, dT&+&V\, dp &\mbox{(Gibbs' potential)}\\
dH&=&T\, dS&+&V\, dp &\mbox{(enthalpy),}\end{array}$$
i.e., $E^{13}=G$, $E^{23}=F$, $E^{14}=H$ and $E^{24}=E$.

If we arrange the energy functions in lexicographic order 
i.e. as $E^{13}$, $E^{14}$, $E^{23}$, $E^{24}$ and denote
them by $5$, $6$, $7$ and $8$ in this order, 
then we can incorporate them into our system.
For it follows from the definitions and simple substitutions that
\begin{eqnarray*}
dE^{13}&=& (y-v f_1)dx - f_2 v dy\\ 
dE^{14}&=&(u f_1+y) dx + u f_2 dy\\
dE^{23}&=&-v f_1 dx +(x-vf_2) dy\\
dE^{24}&=&uf_1 dx+(uf_2-x) dy
\end{eqnarray*}
and so
$$\begin{array}{cclcccl}
(5,1,2)& =& y-gf_1, && (5,2,1)&=& -gf_2\\
(6,1,2)&=&y+fg_1,&& (6,2,1)&=&fg_2\\
(7,1,2)&=&-gf_1,&& (7,2,1)&=&-x-gf_2\\
(8,1,2)&=&fg_1,&& (8,2,1)&=&-x+fg_2.
\end{array}$$

One of the potentially irritating features of the thermodynamical identities is
that many are related by a simple swapping of the variables while this
is accompanied by changes of sign which seem at first sight 
to be random.
The simplest example is displayed by the four Maxwell relations.
We can systematise such computations by 
introducing the symbol $[a,b;c,d]$ for the Jacobi determinant of the mapping
$(c,d)\mapsto (a,b)$ i.e.
 $$[a, b;c, d] = (a, c, d)(b, d, c) - (a, d, c)(b, c, d).$$
The determinant then takes care of the sign.  

For example $[3,4;1,2]$ is the Jacobian
$\dfrac{\partial(u,v)}{\partial(x,y)}$ and so is $1$
(in this case this denotes the number $1$),
$[3, 2;4, 1]$ is $\dfrac{\partial(u,y)}{\partial(v,x)}$
and so is $=-\dfrac{f_1}{g_2}=-\dfrac{(3, 1, 2)}{(4, 2, 1)}$.
\noindent
Note that there are $1,680$ such Jacobians.  
However, lest the reader despair,
we will display below simple rules 
which allow us to express them all in terms of our primitive quantities ($f$
and $g$ together with their partials and, of course, $x$ and $y$).
Later, we append a Mathematica notebook which computes all such expressions
at the press of a button.
\section{Higher derivatives}

Some of the thermodynamical identities involve higher derivatives and
we indicate briefly how to incorporate these into our scheme.
We use the self-explanatory notation $((a,b,c),d,e)$ for second
derivatives.  Thus $((3,1,2),2,1)$ is just $f_{12}$.  Note that this 
notation allows for such derivatives as 
$\left(\dfrac{\partial}{\partial T} \left(\dfrac{\partial E}{\partial p}
\right )_V\right )_S $ which is $((8,1,2),3,4)$.  Once again, we can express
all such derivatives (there are now 18,816 of them) in terms of  $x$, $y$, $f$,
$g$ and their partials (now up to the second order) using the chain rule.
For
$$((a,b,c),i,j)=((a,b,c),1,2)(1,i,j)+((a,b,c),2,1)(2,i,j) $$
and 
$(a,b,c)$ resp. $(1.i,j)$ and $(2,i,j)$ can be dealt with using the
above tables.

We display below a Mathematica notebook which computes
all expressions of the form $((a,b,c),d,e)$. 

\section{Derived quantities and thermodynamical \\
identities}

The reason why there is a plethora of thermodynamical
identities is simple.  A large number of significant (and also insignificant)
quantities can be expressed or defined as simple algebraic combinations
of a very few (our primitive quantities $x$, $y$, $f$, $g$ and their partials).  Hence there are bound to be many relationships between them.
Our strategy to verify (or falsify) an identity is to use the above 
methods to express both sides 
in terms of  these quantities and check whether 
they agree.

Of course, there are myriads of such quantities and identities
and we can only bring a sample.  Thus we have
$$c_V =T \left(\frac{\partial S}{\partial T} \right )_V $$
{\it the heat capacity at constant volume} and
$$c_p = T
\left(\frac{\partial S}{\partial T} \right )_p $$
{\it the heat capacity at constant pressure}. In our formalism,
$c_V=f (4,3,2)$ and $c_p=f(4,3,1)$ and so, from our tables,
$$c_V=f\dfrac{g_1}{f_1}, \quad c_p=f\dfrac{g_2}{f_2} .$$

Hence for the important quantities
$\gamma =\dfrac {c_p}{c_V}$ and ${c_p}-{c_V}$
we have  $\gamma =\dfrac {f_2 g_1}{f_1 g_2}$ and 
${c_p}-{c_V}=f\dfrac1{f_1 f_2}.$

We illustrate our method  by verifying the simple identity:
$$c_p-c_V=T\left(\dfrac{\partial P}{\partial T} \right)_V
\left(\dfrac{\partial V}{\partial T} \right)_p .$$
Using the tables above, we can easily compute both sides in terms of our 
primitive expressions
and get $\dfrac f{f_1f_2}$ in each case. 
\section{Computing $(a,b,c)$ and $((a,b,c),d,e)$}
We can summarise these results in the following formulae:
$$[a,b;c,d]=\frac{(a , 1,2 )( b,2 ,1 )-( a,2 ,1 )(b ,1 ,2 )}{(c ,1 ,2 )(d ,2 ,1 )-(c,2 ,1 )(d ,1,2)}
$$
and so
$$(a,b,c) =[a,c;b,c]=\frac{(a , 1,2 )( c,2 ,1 )-( a,2 ,1 )(c ,1 ,2 )}{(b ,1 ,2 )(c ,2 ,1 )-(b,2 ,1 )(c,1,2)}$$
which allow us to systematically compute any of the derivatives of the form 
$(a,b,c)$ in terms of our primitives $x$, $y$, $f$, $g$, $f_1$, $f_2$, $g_1$ 
and $g_2$.

For the second derivatives we substitute $$\phi=\dfrac{(a , 1,2 )( c,2 ,1 )-( a,2 ,1 )(c ,1 ,2 )}{(b ,1 ,2 )(c ,2 ,1 )-(b,2 ,1 )(c ,1,2)}$$
into the formula

$$(\phi,d,e)=\frac{(\phi , 1,2 )(d ,2 ,1 )-( \phi,2 ,1 )(d ,1 ,2 )}{(d,1 ,2 )(e ,2 ,1 )-(d,2 ,1 )(e ,1,2)}
$$
to compute $((a,b,c),d,e)$ in terms of our primitive terms (this time with the 
first and second derivatives of $f$ and $g$).
The advantage of these formulae is, of course, that one can write 
a simple programme to compute them.
(It is always tacitly assumed in the above formula 
that the appropriate conditions which allows 
a use of the inverse function theorem hold).

The formulae developed here suffice to create Mathematica notebooks to
calculate any of the above quanitities both for the general case
and for specific gas models and we display these below.
\section{A notational survival kit}

We emphasise that the numerical code for the  various thermodynamical
quantities is a mere construct to facilitate their computation
(ideally with the aid of suitable software) and the final goal is to
express them all in terms of the basic quantities ($x$, $y$, $f$, $g$
and the partials of the latter).  It is then a routine matter to
translate these into the standard terminology of thermodynamics
if so required. 
For the convenience of the reader we give a dictionary of the relationships 
between them:
$$\begin{matrix}
1&2&3&4&5&6&7&8\\
p&V&T&S&\Phi&W&F&E\\
x&y&u&v&E^{13}&E^{14}&E^{23}&E^{24}.
\end{matrix}$$
Of course, $p$ is pressure, $V$ volume, $T$ temperature, $S$ entropy
and $\Phi$, $W$, $F$ and $E$ are free enthalpy, enthalpy, free energy
and energy respectively.

\newpage
\section{Mathematica notebooks}
\noindent 
\subsection{A notebook for 
$(a,b,c)$, $[a,b;c,d]$ and $((a,b,c),d,e)$}
We start with a notebook which computes $(a,b,c)$, $([a,b;c,d]$ 
and $((a,b,c),d,e)$ for any combination
of $a$, $b$ and $c$ between $1$ and $8$. 

\begin{verbatim}
Clear[ff1, ff2, ff3, ff4, ff5, ff, gg]; 
ff1[a_, 1, 2] = 
 If[a == 1, 1, 
  If[a == 2, 0, 
   If[a == 3, D[ff[x, y], x], 
    If[a == 4, D[gg[x, y], x], 
     If[a == 5, y - gg[x, y] D[ ff[x, y], x], 
      If[a == 6, y + D[gg[x, y], x] ff[x, y], 
       If[a == 7, -gg[x, y] D[ff[x, y], x], 
        If[a == 8, D[gg[x, y], x] ff[x, y]]]]]]]]];

ff2[a_, 2, 1] = 
  If[a == 1, 0, 
   If[a == 2, 1, 
    If[a == 3, D[ff[x, y], y], 
     If[a == 4, D[gg[x, y], y], 
      If[a == 5, -gg[x, y] D [ff[x, y], y], 
       If[a == 6, D[gg[x, y], y] ff[x, y], 
        If[a == 7, -x - gg[x, y] D[ff[x, y], y], 
         If[a == 8, -x + ff[x, y] D[ gg[x, y], y]]]]]]]]];
ff4[a_, b_, c_, 
   d_] = (ff1[a, 1, 2] ff2[b, 2, 1] - 
     ff2[a, 2, 1] ff1[b, 1, 2])/(ff1[c, 1, 2] ff2[d, 2, 1] - 
     ff2[c, 2, 1] ff1[d, 1, 2]);
ff3[a_, b_, c_] = ff4[a, c, b, c];
ff5[a_, b_, c_, d_, 
   e_] = (D[ff3[a, b, c], x] ff2[d, 2, 1] - 
     D[ff3[a, b, c], y] ff1[d, 1, 2])/(ff1[d, 1, 2] ff2[e, 2, 1] - 
     ff2[d, 2, 1] ff1[d, 1, 2]);


\end{verbatim}

Here we have defined three functions ff$3$, ff$4$ and ff$5$
of $3$, $4$ and $5$ arguments, the variables having whole number
values between $1$ and $8$.  The inputs $a,b,c$ , $a,b,c,d$ or 
$a,b,c,d,e$ yield the values $(a,b,c)$, $[a,b;c,d]$ and $((a,b,c),d,e)$ 
respectively.  In order to show that
it can cope with the most complicated gas models we have used the following
one which combines the Van der Waals gas and the Feynman gas 
(see [Co] for details).

\subsection{A synthesis} 
A number of models of real gases can be subsumed in the following one.
$$u = \phi\left( \left(x+\frac a{y^2}\right)(y-b)\right) \qquad 
v= \ln\left(\left(x+\frac a{y^2}\right)(v-b)^{u(x, y)}\right).$$
$\gamma$, the adiabatic exponent, is a function of one variable and
$\phi$ is a   primitive of $\dfrac1{\gamma-1}$.
This model has the double advantage that it incorporates the 
van der Waals gas and also one introduced by Feynman which allows for the 
fact that in real gases, $\gamma$ depends on temperature.
If the constants $a$ and $b$ vanish, we get what we call the Feynman gas
(this model appears in [Fe]---see [Co] for a more detailed description).
We get the van der Waals gas when $\gamma$ is constant and this specialises to
the ideal gas when $a$ and $b$  vanish.

We can now compute any of the above expressions for this model, using the
above notebook.

\subsection{Thermodynamics identities and 
Gr\"obner bases}
We can proceed further along the path to automatisation by invoking the concept and methods of Gr\"obner bases.  Rather than give a systematic treatment,
we  again display
a corresponding notebook.

The input  below consists of two list of symbols which
use an obvious code---$x412$, for example, denotes the partial 
derivative $(4,1,2)$.
The second list is of the quantities which occur in the formulae we are 
interested in.  They are regarded as variables and our setting is the
ring of polynomials in these variables.
This list can be chosen at the discretion of the user (there are too
many candidates to include all of them in one notebook).  
The first list is  of the relationships between these symbols (each 
expression is to be thought of as being set equal to zero) beginning 
with the Maxwell relation
followed by the formulae deduced above. 
In order to keep things simple, we have only used the relationships
which we computed above.  They can, of course, be enriched by triples which
were computed by the above Mathematica programme.

\begin{verbatim}
Clear[p, q]; p = {x312 x421 - x321 x412 - 1, x134 - x421, x143 + x321,
   x234 + x412, x243 - x312, x421 x341 - x321, x314 x421 - 1, 
  x241 x421 - 1, x214  x421 + x412, x342 x412 - x312, x324 x412 + 1, 
  x412 x142 - 1, x124 x412 + x421, x432 x312 - x412, x423 x312 - 1, 
  x132 x312 - 1, x123 x312 + x321, x431 x321 - x421, x413 x321 + 1, 
  x231 x321 - 1, x213 x321 + x312, x612 - x2 - x3  x412, 
  x621 - x3 x421, x512 - x2 + x4  x312, x521 + x4 x321, 
  x712 + x4  x312, x721 + x1 + x4  x321, x812 - x3  x412, 
  x821 + x1 - x3 x421};
q = {x1, x2, x3, x4, x112, x121, x212, x221, x312, x321, x412, x421, 
   x134, x143, x234, x243, x341, x314, x241, x214, x342, x324, x142, 
   x124, x432, x423, x132, x123, x431, x413, x231, x123, x512, x521, 
   x612, x621, x712, x721, x812, x821};
GroebnerBasis[p, q]
\end{verbatim}

The output  below is a (Gr\"obner) basis for the ideal generated by the
above list of relationships in our polynomial ring.
Hence every entry is, when set equal to zero, a (new) identity.
\begin{verbatim}

{x213 x621 + x712 + x213 x721 - x213 x821, x521 - x621 - x721 + x821, 
 x512 - x612 - x712 + x812, 
 x231 + x413, -x231^2 x621 + x431 x712 + x213 x431 x721 - x431 x812 - 
  x213 x431 x821, -1 + x123 x213, x621 + x123 x712 + x721 - x821, 
 x123 x231^2 x621 + x431 x621 + x123 x431 x812, x132 + x123 x231, 
 x123 x231 + x423, -x123 x231^2 - x431 + x432, x621 + x124 x812, 
 x124 x231^2 - x431 + x124 x213 x431, 
 x123 x124 x231^2 - x123 x431 + x124 x431, 1 - x124 x213 + x142 x231, 
 x124 x231 + x142 x431, x142 + x324, 
 x342 x712 + x213 x342 x721 + x142^2 x213 x812 - x342 x812 - 
  x213 x342 x821, x142 x213 + x231 x342, -x124 x213 + x342 x431, 
 x342 x621 - x142^2 x812 + x123 x342 x812, 
 x142^2 - x123 x342 + x124 x342, 
 x214 x712 + x213 x214 x721 - x213 x812 - x213 x214 x821, 
 x214 x621 + x812, -x231^2 - x213 x431 + x214 x431, -1 + 
  x124 x214, -x142^2 x213 x214 - x213 x342 + x214 x342, 
 x142 x214 + x241, x142 x214 + x314, -x142^2 x214 + x341 - x342, 
 x243 x712 + x213 x243 x721 + x142 x213 x812 - x243 x812 - 
  x213 x243 x821, x213 + x231 x243, 
 x243 x621 - x142 x812 + x123 x243 x812, x142 - x123 x243 + x124 x243,
  x142 x243 - x342, -x142 x213 x214 - x213 x243 + x214 x243, -x231 + 
  x234 + x243 x431, x143 - x123 x243, x134 + x123 x243 x431, 
 x421 + x123 x243 x431, x231 + x412 - x243 x431, 
 x123 x243 + x321, -x243 + x312, 
 x4 + x231 x621 + x231 x721 - x231 x821, x3 - x142 x812, 
 x2 - x612 + x812, x1 - x621 + x821}
\end{verbatim}

\paragraph{The literature:} The method of Gr\"obner bases which is used
in Mathematica is based on the so-called Buchberger algorithm (see [Bu])
and there is an extensive literature about it.

We would also like to mention two works ([Br] and [Ja]) which are dedicated
to the development of systematic approaches to thermodynamical identities
and which were useful in our approach.
\paragraph{Acknowledgements:}
I would like to express my thanks to the following scholars whose input
was very important in chrystalising the ideas presented in this paper:
the late P.A. Samuelson, T. Russell, P.F.X. M\"uller (who pointed out 
the Feynman
model to me) and Elena Kartashova.

\end{document}